\documentclass{elsart}
\usepackage{epsfig}
\begin{document}
\begin{frontmatter}
\title{Supernovae~and~Positron~Annihilation~Radiation}

\vspace{-7mm}
\author[c1]{P.A.~Milne}
\author[c2]{J.D.~Kurfess, R.L.~Kinzer}
\author[c3]{M.D.~Leising}
\address[c1]{NRC/NRL Resident Research Associate, Naval Research Lab,
Code 7650, Washington DC 20375}
\address[c2]{Naval Research Lab, Code 7650, Washington DC 20375}
\address[c3]{Clemson University, Clemson, SC 29631}

\vspace{0mm}
\begin{abstract}

Radioactive nuclei, especially those created in SN explosion, have long 
been suggested to be important contributors of galactic positrons. In this
paper we describe the findings of three independent OSSE/SMM/TGRS studies 
of positron annihilation radiation, demonstrating that the three studies 
are largely in agreement as to the distribution of galactic annihilation 
radiation. We then assess the predicted yields and distributions of 
SN-synthesized radionuclei, determining that they are marginally compatible 
with the findings of the annihilation radiation studies. 

\end{abstract}

\end{frontmatter}

\vspace{-12mm}
\section{OSSE/SMM/TGRS Observations}

\vspace{-5mm}
Nine years of observations made with the Oriented 
Scintillation Spectrometer Experiment (OSSE) on-board NASA's 
COMPTON observatory (1991-2000)\cite{john93}, eight years of observations 
made with the Gamma-Ray Spectrometer on-board the Solar
Maximum Mission (SMM) (1980-1989)\cite{shar88}, and two years 
of observations made with the Transient Gamma-Ray
Spectrometer (TGRS) on-board the WIND mission (1995-1997) \cite{harr98}
 have been utilized to study the galactic 
distribution of positron annihilation radiation. 
The OSSE instrument featured a 3.8$^{\circ}$ x 11.4$^{\circ}$ FWHM 
FoV, a $\sim$3 x 10$^{-5}$ photons cm$^{-2}$ s$^{-1}$ line sensitivity 
(10$^{6}$~s on-source time), and a 45 keV energy resolution at 511 keV. 
These detector attributes have permitted the first detailed studies of the 
distribution of annihilation radiation in the inner radian of the Galaxy. 
The annihilation of positrons with electrons gives rise to two 
spectral features, a line emission at 511 keV and a positronium 
continuum emission (which increases in intensity with energy 
roughly as a power 
law up to 511 keV and falls abruptly to zero above 511 keV)\cite{ore49}. 
The TGRS instrument, which featured a germanium detector with excellent 
energy resolution, has 
demonstrated that the integrated flux from the inner radian is best 
described as a narrow 511 keV line (FWHM~$\leq$~1.8 keV) and a positronium 
continuum to 511 keV line ratio of $\sim$ 3.6 (which corresponds to a 
positronium fraction of f$_{Ps}$=0.94)\cite{harr98}. 

Purcell et al. (1997) \cite{purc97}
reported results from combined 
OSSE/SMM/TGRS studies of the 511 keV line component of 
annihilation radiation. They found the 511 keV emission to be comprised of 
three components; 1) an intense bulge emission, 2) a fainter disk emission, 
and 3) an enhancement of emission at positive latitudes (hereafter called 
a PLE). The PLE was also reported by Cheng et al. (1997)\cite{chen97}, and 
has been interpreted to be an ``annihilation fountain" 
by Dermer \& Skibo\cite{derm97}.
Purcell et al. (1997) characterized the emission via mapping, employing the SVD 
matrix inversion algorithm, and via model fitting, testing the 
combination of a spheroidal Gaussian bulge, a disk that is flat in 
longitude to $\pm$40$^{\circ}$ and Gaussian in latitude (FWHM = 9$^{\circ}$), 
and a spheroidal PLE. The two characterizations differ in the thickness of 
the Gaussian disk (SVD being narrower) and the extension of the PLE. The 
enhancement of the PLE varied 
from 1.5  x 10$^{-4}$ phot cm$^{-2}$ s$^{-1}$ for the SVD map to 9 x 
10$^{-4}$ phot cm$^{-2}$ s$^{-1}$ for the 2D Gauss. PLE (FWHM~=~16.4$^{\circ}$). 
The positron annihilation rates suggested
by these two approaches are (4.2$\pm$0.5) x 10$^{43}$ e$^{+}$~s$^{-1}$ (SVD) and
(3.3$\pm$0.50) x 10$^{43}$ e$^{+}$~s$^{-1}$ (2D Gauss.), with B/D ratios of
0.5 and 0.3.                                 

Parallel studies were performed by Kinzer et al. (2001). 
Those studies investigated line and continuum positron annihilation 
radiation from the galactic plane, finding that the two emissions are 
similarly distributed within the statistical precision of the data.
Kinzer et al. (2001) suggest two spatial distributions to explain the OSSE/SMM 
observations. The first model is comprised of a 2D Gaussian bulge 
(4.9$^{\circ}\pm$0.7$^{\circ}$($b$) x 6.3$^{\circ}\pm$1.5$^{\circ}$($l$)), 
and two disks; a 2D Gaussian disk (12$^{\circ}$($b$) x 
35$^{\circ}\pm$10$^{\circ}$($l$)) and a widely-distributed CO disk (also 
with 12$^{\circ}$ latitudinal thickness). The second model is 
comprised of a slightly ellipsoidal bulge which follows the ``R$^{1/4}$" 
distribution function  and an exponential disk (refer to Kinzer et al. (2001) for 
details of these functions). The positron annihilation rates suggested 
by these models are 3.9$\pm$0.4 x 10$^{43}$ e$^{+}$~s$^{-1}$ and 
3.1$\pm$0.4 x 10$^{43}$ e$^{+}$~s$^{-1}$ respectively, with B/D ratios of 
0.7 and 3.
 
\begin{figure}
\centerline{\epsfig{file=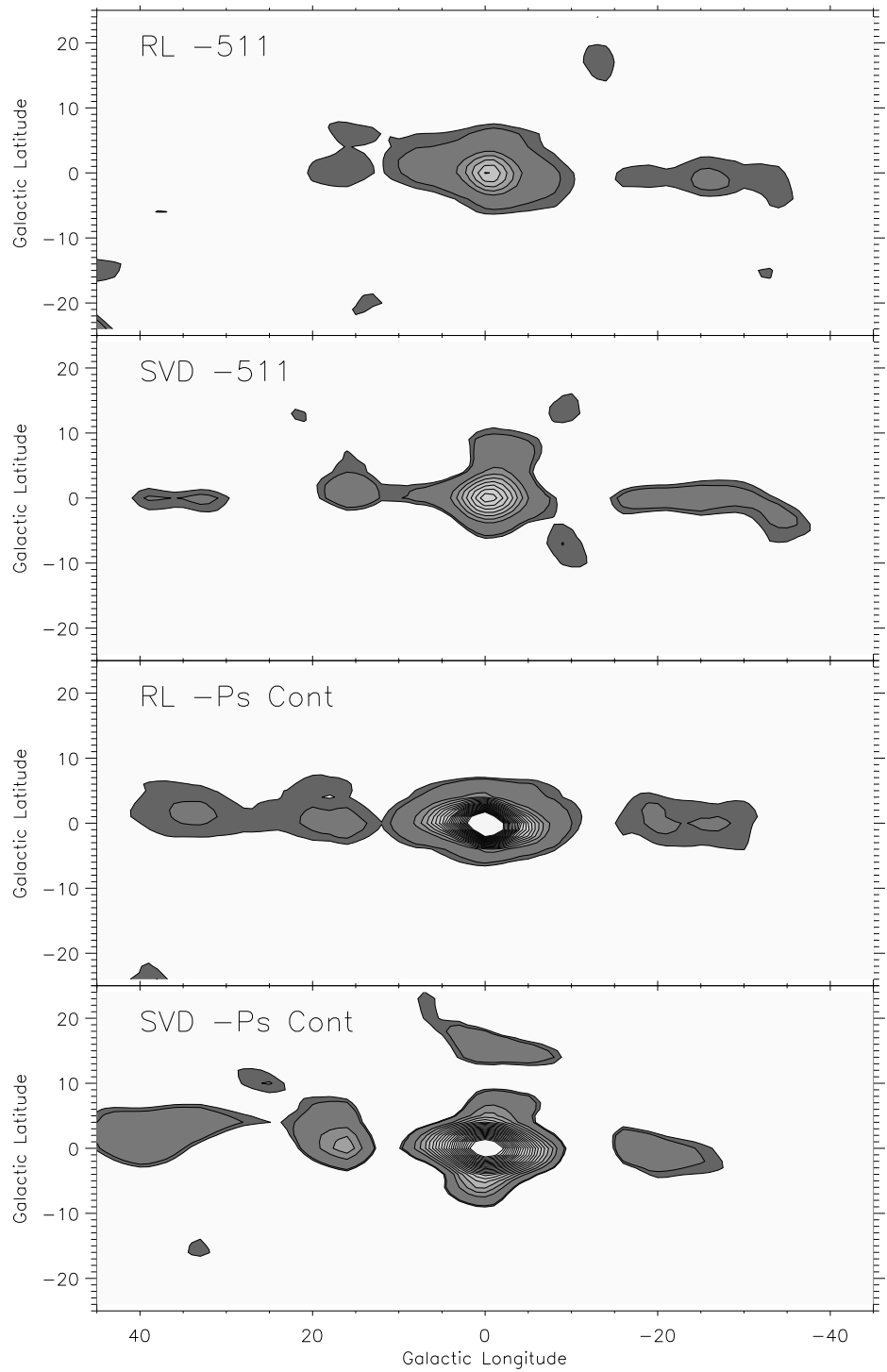, width=13cm}}
\vspace{-.8cm}
\caption{Four characterizations of positron annihilation radiation 
from mapping algorithms. The upper 
two panels are Richardson-Lucy and SVD maps of 511 keV line 
emission. The lower two panels are Richardson-Lucy and SVD maps of 
positronium continuum emission. The SVD mapping algorithm was used 
in Purcell et al. (1997).}
\end{figure}

We report here updates from 
our continuing analysis which extends the study of Purcell et al. (1997)
(see also Milne et al. (1998,1999) \cite{miln98,miln99}). The 
primary differences between current studies and Purcell et al. (1997) 
are; 1) the inclusion of 
more observations, both archival and data collected after 
Purcell et al. (1997), and 2) 
reporting maps of the positronium continuum emission 
in addition to the 511 keV line. To extract the positronium continuum 
component from the underlying galactic continuum emission, we 
widened the spectral modeling to include thermal bremsstrahlung and 
exponentially-truncated power-law models. We also removed high-energy 
diffuse continuum emission following a prescription from Kinzer et al. 
(1999), distributing the emission spectrally according to a  
power-law  ($\alpha$ = -1.65) and spatially according to a 
90$^{\circ}$ x 5$^{\circ}$ 2D Gaussian\cite{kinz99}.

Maps made from mapping algorithms applied to the resulting data-set 
are shown in Figure 1. The top two panels show Richardson-Lucy 
and SVD maps of 511 keV line emission, the lower two panels 
show Richardson-Lucy and SVD maps of positronium continuum emission. 
Although not identical, the four maps are similar in appearance, with 
all exhibiting an intense bulge emission 
and a fainter planar emission.\footnote{We refer the reader to 
Milne et al. 2001 for a discussion of current investigations of 
the PLE.} Pairings 
of bulge and disk components suggest the same families of solutions for 
both the line and positronium continuum emissions. 
Both suggest that the bulge-to-disk ratio can vary from 0.2 -3.3 depending 
upon whether the bulge component features a halo (which leads to a 
large B/D). All of these comparisons support the Kinzer et al. (2001) suggestion that 
the two annihilation components are similarly distributed. 
Two bulge-disk combinations that span the range of 
favored solutions are shown in Figure 2. The upper  
panel shows a ``bulge-dominated" solution, where a halo bulge plus a 
thin disk combine to annihilate 4.1 x 10$^{43}$ e$^{+}$ s$^{-1}$ with a 
B/D = 3.3. 
The lower panel 
shows a ``disk-dominated" solution, where a 2D Gaussian bulge (without 
a halo) plus a thick disk combine to annihilate
4.1 x 10$^{43}$ e$^{+}$~s$^{-1}$ (B/D =0.2).

\begin{figure}
\centerline{\epsfig{file=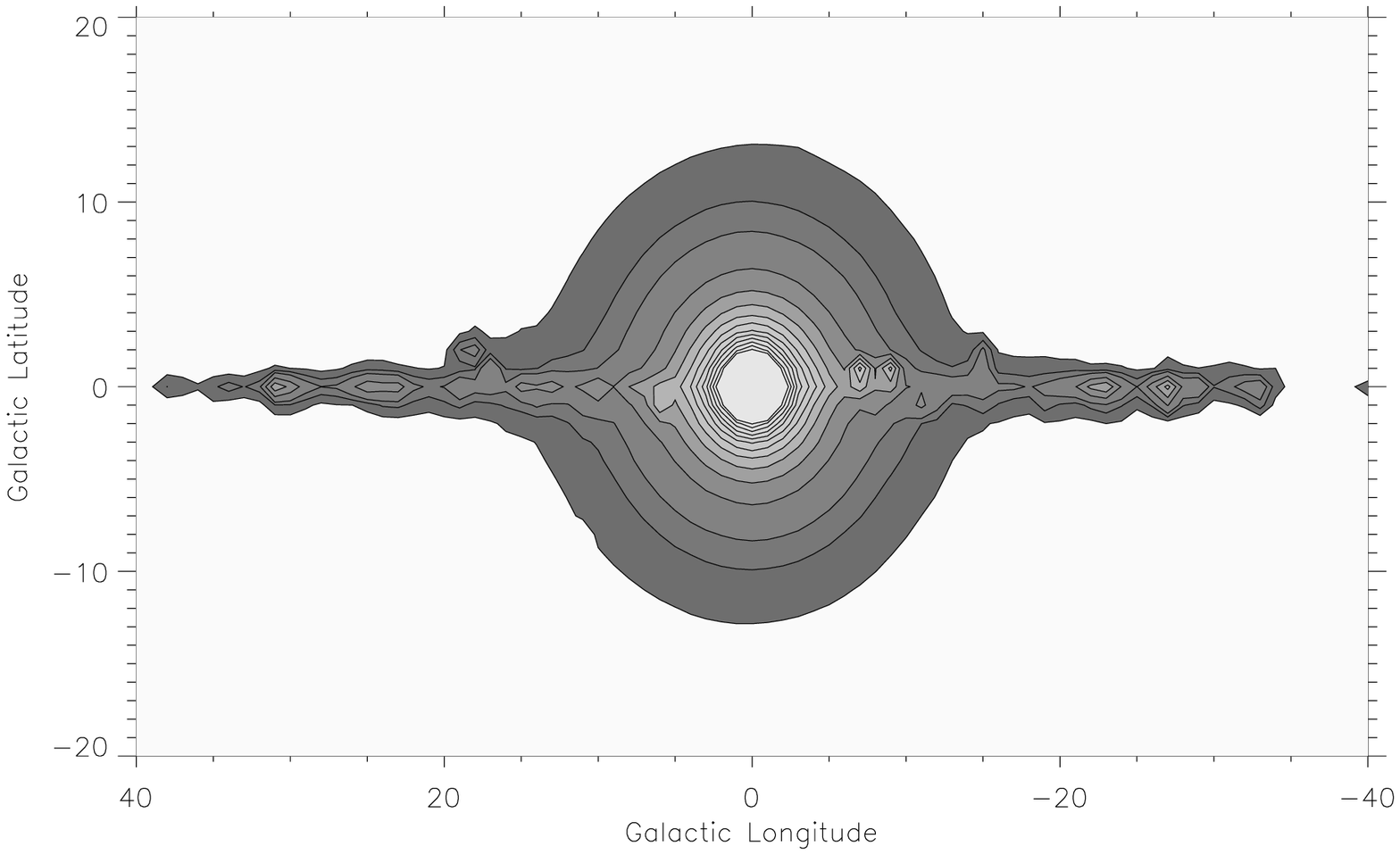, width=8.4cm}}
\centerline{\epsfig{file=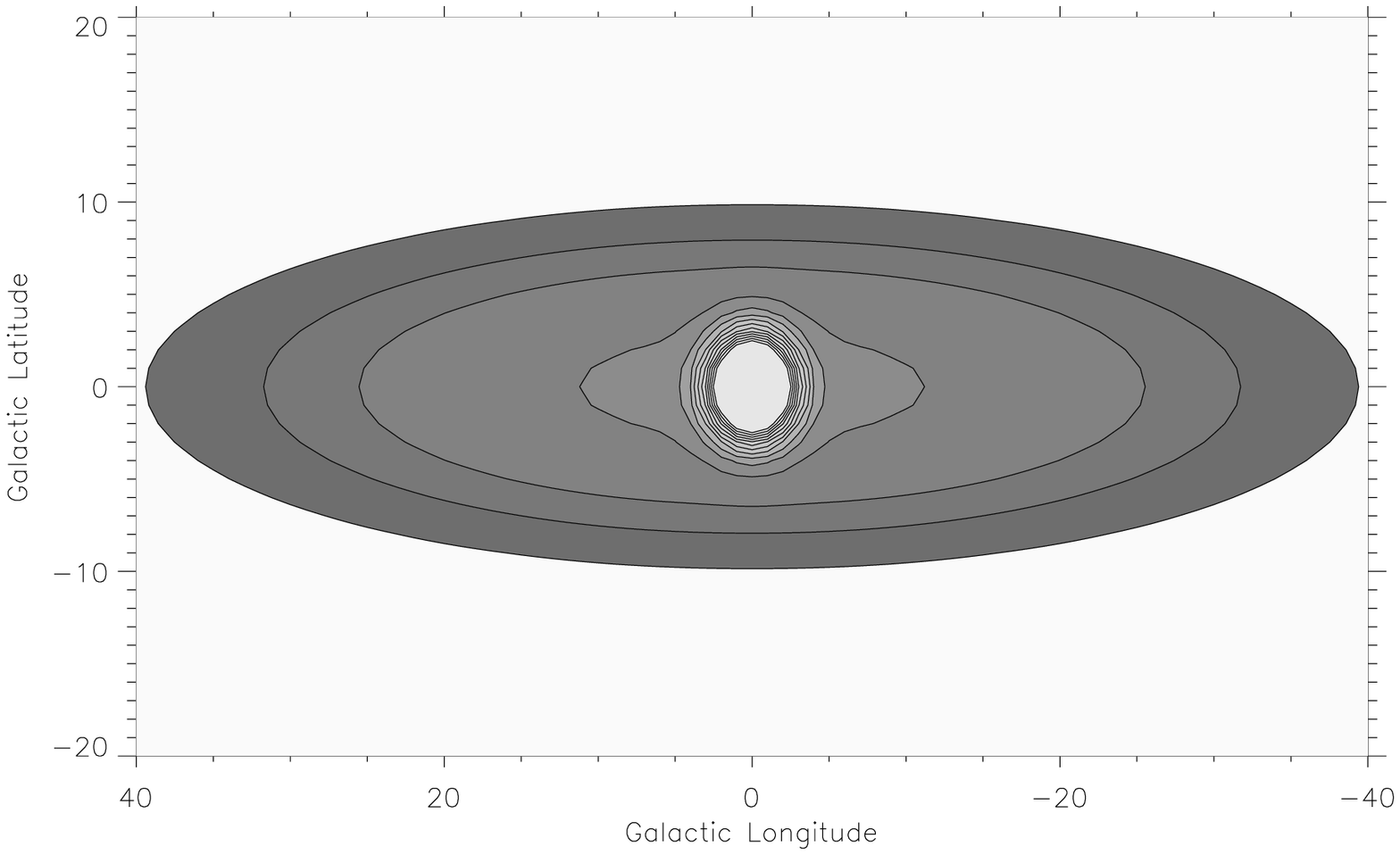, width=8.4cm}} 
\vspace{-.3cm}
\caption{Two characterizations of positron annihilation radiation 
from model-fitting. 
The upper panel shows a ``bulge-dominated" solution.
The lower panel shows a ``disk-dominated" solution. The 
``bulge-dominated" solution is similar to the nova-spheriod model of 
Kinzer et al. (2001), the ``disk-dominated" solution is similar to the three 2D 
Gaussian model of Purcell et al. (1997) and to the two 2D 
Gaussian + CO model of Kinzer et al. (2001).}
\end{figure}

\vspace{-7mm}
\section{Supernovae and Galactic Positrons}

\vspace{-6mm}
The B/D ratios favored by the current work bracket 
those found by Purcell et al. (1997) and Kinzer et al. (2001), 
allowing a range of 0.2 $\leq$ B/D 
$\leq$ 3.3. The positron annihilation rate is suggested to lie within 
the range of (3.1 -4.2) x 10$^{43}$ e$^{+}$ s$^{-1}$. 
To compare suggested SN positron production rates with 
observations, a number of {\em ad-hoc} assumptions must be made. First, 
we assume that positrons do not travel large distances from their 
site of origin, so that source B/D ratios are assumed to equal 
observed annihilation radiation B/D ratios. Second, we assume that there 
is no leakage of positrons from the Galaxy nor enrichment of positrons 
from extra-galactic sources. Third, we assume that the number of 
galactic positrons is not changing, that there is a steady-state 
population of positrons. These three assumptions allow us to 
compare annihilation radiation fluxes with positron production rates.  
Fourth, we assume that the SNe in the Galaxy scale with B-band 
luminosity the same way as SNe in other late spiral galaxies. 
This permits the use of extra-galactic SN rates as an indicator of the 
recent SN activity in the Galaxy. 

SNe Ia produce positrons primarily from the decay of 
$^{56}$Ni $\rightarrow$ $^{56}$Co  $\rightarrow$ $^{56}$Fe, although 
the decay of $^{44}$Ti $\rightarrow$ $^{44}$Sc  $\rightarrow$ $^{44}$Ca 
also contributes positrons. Chan \& Lingenfelter (1993) simulated 
positron transport through SN Ia models and suggested that $\sim$ 5\% 
of $^{56}$Co-decay positrons escape the ejecta for favorable magnetic 
field scenarios. Milne et al. (1999,2001a) followed up that study by 
comparing SN Ia light curves with simulations. That study found that the 
light curves suggest that $\sim$8 x 10$^{52}$ positrons escape the 
ejecta from a normal type Ia supernova. Assuming the typical $^{44}$Ti 
yield in a SN Ia to be  2 x 10$^{-5}$ M$_{\odot}$, and 100\% escape 
of those positrons, only 5 x 10$^{50}$ $^{44}$Ti positrons would 
be produced in a single SN Ia. Assuming the SN Ia rate in the Galaxy to 
be 0.4 SN Ia per century, this translates to 1.0 x 10$^{43}$ 
and 6  x 10$^{40}$ positrons s$^{-1}$ from $^{56}$Co and $^{44}$Ti decays. 
The SN Ia contribution to galactic positrons principally depends 
upon whether positrons escape in quantity from the SN ejecta.

Estimating the massive star (SNII/Ib and WR star) contribution to galactic positrons is
more complicated.
They produce positrons through the decays of
 $^{44}$Ti $\rightarrow$ $^{44}$Sc  $\rightarrow$ $^{44}$Ca, and
$^{26}$Al $\rightarrow$ $^{26}$Mg, the latter from both hydrostatic and explosive phases.  The
predicted yields of these
isotopes vary dramatically in individual events.
Woosley \& Weaver (1995) suggest
that they range from $\leq$~10$^{-6}$ up to 2.3 x 10$^{-4}$
~M$_{\odot}$ for $^{44}$Ti, and 1.7 x 10$^{-5}$
--3.6 x 10$^{-4}$~M$_{\odot}$ for
$^{26}$Al\cite{woos95}.
SN II/Ib rates have been reported to range between
1.5 --3.3 SNe II/Ib per century\cite{capp00,hata97}.
To help to constrain our estimates, we use observations
of 1.809 MeV line emission, which is produced in 100\% of $^{26}$Al decays.
These suggest about 3 x 10$^{42}$ $^{26}$Al decays s$^{-1}$ occur entirely
in the galactic disk (see Pluschke in these proceedings) supplying 2.4 x 10$^{42}$ positrons
s$^{-1}$.

The positron contribution of $^{44}$Ti is best constrained by galactic chemical evolution and the
solar abundance of the daughter $^{44}$Ca. This requires that 2--4 x10$^{-6}$ M$_{\odot}$ of
$^{44}$Ti are now ejected into the galactic disk \cite{timm96}. We note, however, that the production
obtained by multiplying the above rates and yields falls short of this. The implied positron production
rate is 1.6--3.2 x 10$^{42}$ s$^{-1}$, and we estimate the total massive star contribution to the disk
to be 4--6 x 10$^{42}$ positrons s$^{-1}$.

Combining these estimates, 30\% -50\% of galactic positrons may be
explained by thermonuclear SNe and massive stars. This appears
promising, but this model must also agree with the observed B/D ratio of
annihilation radiation. SNe Ia occur in both bulges and disks of spiral
galaxies, with suggested B/D ratios ranging from 1/2 to 1/7
\cite{howe00,hata97}. Our
 estimates suggest that the B/D ratio ranges from 0.1 to 0.6. These values are
at the low extreme of the fitted  B/D ratios from the annihilation data (which are
realized when halo-less bulges are combined with thick disks).
If $^{56}$Co positrons {\bf do not} escape from the ejecta of SNe Ia,
in contrast to the studies of Milne et al. (1999,2001) , then the
massive star contribution is at most 20\% of galactic positrons, and
the bulge component is entirely due to a different type of source.

\vspace{-6mm}
\section{Discussion}

\vspace{-3mm}
After accepting a couple of {\em ad-hoc} assumptions, 
we have shown that SNe are likely to be a dominant contributor to 
galactic positrons. We find that the maximum SN + $^{26}$Al contribution 
occurs with low B/D ratios, requiring that the remaining 45\% of 
positrons be produced by a wholly-bulge source type. Lowering the 
SN II/Ib contribution somewhat relaxes this requirement, but the 
remaining sources must collectively have a large B/D ratio. Potential 
sources include; 1) a recent galactic center starburst that produced a 
population of bulge positrons through massive star nucleosynthesis 
and SNe II, 2) positrons produced by a galactic center compact source, 
3) nucleosynthesis in novae. All three of these sources are contrained 
by theory or observations. The starburst is constrained by the absence 
of a bulge component in the 1.809 MeV maps and the failure to detect 
1.173 MeV emission from the decay of SN-produced $^{60}$Fe. OSSE 
observations of 511 keV line emission from 
the galactic center limit the contribution of a compact source to less 
than 1.5 x 10$^{-4}$ photons cm$^{-2}$~s$^{-1}$\cite{miln01a}. This 
is less than 6\% of the total annihilation radiation. However, if the 
positrons diffuse throughout the bulge, a GC compact source 
could contribute at a higher level. Novae production of positrons from 
radionuclei (other than $^{26}$Al) are constrained by simulations 
that suggest that few positrons produced in CO novae would 
escape \cite{leis87}, and low $^{22}$Na yields in ONe novae\cite{jose99}. 

There are too many uncertainties involved in these comparisons to 
determine the validity of any of the assumptions that were used. 
Advances in the studies of the gamma-ray and optical emission from 
SNe will eventually resolve these questions. At present, we must be 
content with interpreting hints as to the SN contribution to positrons 
in the Galaxy.


\vspace{-6mm}
\begin{thebibliography}{}      

\vspace{-3mm}
\bibitem{john93} Johnson, W.N., et al. {\it ApJS}, {\bf 86}, 693 (1993).
\bibitem{shar88} Share, G.H., et al., {\it ApJ} {\bf326}, 717 (1988).
\bibitem{harr98} Harris, M.J., et al., {\it ApJ} {\bf 501}, L55 (1998).
\bibitem{ore49} Ore, A. \& Powell, J.L., {\it Phys Rev}, {\bf 75}, 11 (1949).
\bibitem{purc97} Purcell, W.R., et al., {\it ApJ} {\bf 491}, 725 (1997).
\bibitem{chen97} Cheng, L.-X., et al., {\it ApJ} {\bf 481}, L43 (1997).
\bibitem{derm97} Dermer, C.D. \& Skibo, J.G., {\it ApJ} {\bf 487}, L57 (1997).
\bibitem{kinz01} Kinzer, R.L., et al., {\it ApJ}, {559}, 705 (2001).
\bibitem{miln98}Milne, P.A. et al., {\it Astro. Lett. \& Comm.} {\bf 38},
441 (1998).
\bibitem{miln01a}Milne, P.A. et al., {\it in Proceedings of the 6th 
Compton Symposium}, in press (astro-ph 0106157) (2001a).
\bibitem{kinz99} Kinzer, R.L., Purcell W.R., \& Kurfess, J.D., {\it ApJ} {\bf 515}, 215 (1999).
\bibitem{chan93}Chan, K.-W. \& Lingenfelter, R., {\it ApJ}, {\bf 405}, 614 
(1993).
\bibitem{miln99}Milne, P.A., The, L.S., Leising, M.D., {\it ApJS}, {\bf 124}, 
503 (1999).
\bibitem{miln01b}Milne, P.A., The, L.S., Leising, M.D., {\it ApJ-accepted}, 
 (astro-ph 0104185) (2001b).
\bibitem{capp00}Cappellaro, E., Turatto, M., {\it astro-ph}, 0012455 (2000).    
%\bibitem{knod99}Knodlseder, J., {New Astronomy Rev.}, {\bf 44:4}, 315 (2000). 
\bibitem{woos95}Woosley, S.E., Weaver, T.A., {\it ApJS}, {\bf 101}, 181 (1995).
\bibitem{timm96}Timmes, F.~X., Woosley, S.~E., Hartmann, D.~H., \& Hoffman, R.~D., 
{\it ApJ}, {\bf 464}, 332 (1996).
\bibitem{vink01}Vink, J., et al. {\it astro-ph}, 0107468 (2001).
\bibitem{maho84}Mahoney, W., Ling, J.C., Jacobson, A., {\it ApJ}, {\bf 286}, 
578 (1984).
\bibitem{howe00}Howell, A., Wang, L., Wheeler, J.C., {\it ApJ}, {\bf 530},
166 (2000).
\bibitem{hata97}Hatano, K., Fisher, A., Branch, D., {\it MNRAS}, {\bf 290}, 
360 (1997).
\bibitem{leis87}Leising, M.D., \&
 Clayton, D.D., {\it ApJ}, {\bf 323}, 159 (1987).
\bibitem{jose99}Jose, J., Coc, A., \& Hernanz, M., {\it ApJ}, {\bf 520}, 
347 (1999).

\end{thebibliography}
\end{document}